\documentclass[12pt]{article}

\usepackage{amssymb,amsmath}

\usepackage{graphicx}
\usepackage{color}
\usepackage[colorlinks=true
,urlcolor=blue
,citecolor=blue
,linkcolor=blue
,pagecolor=blue
,linktocpage=true
,pdfproducer=medialab
]{hyperref}
\usepackage[a4paper,width=15.5cm]{geometry}
\usepackage{subfigure}

\makeatletter \renewcommand{\@dotsep}{10000} \makeatother

\def\tev{\,{\rm TeV}}

\newcommand{\beq}{\begin{equation}}
\newcommand{\eeq}{\end{equation}}
\newcommand{\bea}{\begin{eqnarray}}
\newcommand{\eea}{\end{eqnarray}}

\begin{document}

\begin{center}

 {\Large

Non-Universal Gaugino Masses and  Natural Supersymmetry

 } \vspace{1cm}

{\large   Ilia Gogoladze\footnote{E-mail: ilia@bartol.udel.edu\\
\hspace*{0.5cm} On  leave of absence from: Andronikashvili Institute
of Physics, 0177 Tbilisi, Georgia.},    \vspace{.3cm} Fariha Nasir\footnote {E-mail: fariha@udel.edu }  and  Qaisar Shafi\footnote{ E-mail:
shafi@bartol.udel.edu} } \vspace{.9cm}

{\baselineskip 20pt \it
Bartol Research Institute, Department of Physics and Astronomy, \\
University of Delaware, Newark, DE 19716, USA  } \vspace{.5cm}

\vspace{1.5cm}
 {\bf Abstract}
\end{center}
We demonstrate that natural supersymmetry  is readily realized in the framework of
$SU(4)_c \times SU(2)_L \times SU(2)_R$  with non-universal gaugino masses. Focusing on ameliorating the little hierarchy problem, we explore the parameter space of this model which yields small fine-tuning measuring parameters (natural supersymmetry) at the electroweak scale ($\Delta_{EW}$) as well as at high scale ($\Delta_{HS}$). It is possible to have both $\Delta_{EW}$ and $\Delta_{HS}$ less than 100 in these models, (2\% or better fine-tuning), while keeping the light CP-even (Standard Model-like) Higgs mass in the 123 GeV-127 GeV range.
The light stop quark mass lies in the range 700 GeV $<m_{\tilde{t}_{1}}<$ 1500 GeV, and the range for the light stau lepton mass is  900 GeV $<m_{\tilde{\tau}_{1}}<$ 1300 GeV. The first two family squarks are in the mass range 3000 GeV $<m_{\tilde{t}_{1}}<$ 4500 GeV, and for the gluino we find 2500 GeV $<m_{\tilde{g}_{1}}<$ 3500 GeV.
We do not find any solution with natural supersymmetry
which yields
significant enhancement for  Higgs production and decay in the diphoton channel.

\newpage

\renewcommand{\thefootnote}{\arabic{footnote}}
\setcounter{footnote}{0}



\section{Introduction  \label{intro}}

The ATLAS and CMS collaborations at the Large Hadron Collider (LHC)  have independently reported the discovery \cite{:2012gk, :2012gu} of a particle with production and decay modes that appear more or less consistent with the Standard Model (SM) Higgs boson of mass $m_h \sim 125$ GeV.
The  Minimal Supersymmetric Standard Model (MSSM)   can accommodate values of $m_h \sim 125 \rm \ GeV$, but this  requires either a very large, ${\cal O} (\mathrm{few}-10)$ TeV,  stop quark mass \cite{Ajaib:2012vc}, or a large soft supersymmetry  breaking (SSB) trilinear $A_t$-term, with a stop quark mass of around a TeV \cite{Djouadi:2005gj}. Such a heavy stop quark leads to the so-called  ``little hierarchy"  (or ``natural supersymmetry") problem \cite{b5} because, in implementing radiative electroweak symmetry breaking, TeV scale quantities must conspire to yield the (100 GeV) electroweak mass scale ($M_{EW}$).  Discussion of natural supersymmetry in light of a 125 GeV Higgs boson  are found in Ref. \cite{Hall:2011aa, Baer:2012mv,Younkin:2012ui}.

In ref. \cite{Baer:2012mv} it was argued  that
in order to satisfactorily resolve the little hierarchy problem in a supersymmetric theory, one should investigate  the fine-tuning condition not only at $M_{EW}$ scale but also at some high, presumably GUT scale, too.  It was found  \cite{Baer:2012mv} that in the Constrained MSSM (CMSSM) case, the
GUT scale fine-tuning condition for the little hierarchy problem is at least ten times more stringent compared to the same condition at  $M_{EW}$. In this paper we adopt their procedure for analyzing the naturalness problem in our class of supersymmetric models.

It has been shown in
\cite{Younkin:2012ui,Abe:2007kf,Gogoladze:2009bd} that non-universal gaugino masses at  $M_{GUT}$
 can help resolve the little hierarchy problem. To show this,
the authors have studied a variety of gaugino mass ratios
obtained from some underlying theories.
In this paper we revisit the study performed in  ref. \cite{Gogoladze:2009bd}
based on the  $SU(4)_c \times SU(2)_L \times SU(2)_R$ (4-2-2) gauge supersymmetry \cite{Pati:1974yy}, which
provides a natural setup for non-universal gaugino masses.
In this paper, we perform a more elaborate study by employing the ISAJET 7.84 package \cite{ISAJET} which can be used to implement the various phenomenological and cosmological constraints. We show that
the little hierarchy problem  can be largely resolved if the $SU(2)_L$ and $SU(3)_c$
gaugino masses satisfy the asymptotic relation $M_2/M_3\approx 3.3$.

In addition to the Higgs discovery, the ATLAS and CMS experiments have reported an excess in Higgs production and decay in the diphoton channel, around $1.4-2$ times larger than the SM expectations. The statistical significance of this apparent deviation from the SM prediction is at present not sufficiently strong to draw a definite conclusion, but if confirmed in the future, it will be clear indication of new physics around the electroweak scale.
The charged superparticles in the MSSM can give a sizable
contribution to the Higgs coupling to photons provided they are sufficiently light. A light third
generation sfermions with large left-right mixing   have been considered to achieve enhancement in Higgs production and decay in the diphoton channel, relative to the SM \cite{Carena:2011aa,Ajaib:2012eb}. In this present paper we estimate the diphoton rate $R_{\gamma \gamma}$  for the parameter space which provides a resolution of the little hierarchy problem.

The layout of this paper is as follows. In  Section \ref{lhp} we briefly summarize the little hierarchy problem in the MSSM.  In Section \ref{ft} we briefly describe the fine-tuning conditions at low and high scales. The $SU(4)_c \times SU(2)_L \times SU(2)_R$ model   we study in this paper is described in Section \ref{model}.
Section \ref{constraintsSection} encapsulates    the scanning procedure and the experimental constraints that  we employ.
Our results are discussed in  Section \ref{results} and
our conclusion are presented in Section \ref{conclusions}.

\section{Little Hierarchy Problem in the MSSM  \label{lhp}}

At tree level the lightest CP-even (SM-like) Higgs boson mass $m_h$ in the MSSM is bounded from above
by the mass of the $Z$ boson
\begin{equation}
m_h < M_Z.
\label{ee1}
\end{equation}
It is clear from Eq.(\ref{ee1})  that significant radiative corrections are needed in order to accommodate values of $m_h \sim 125 \rm \ GeV$.
 For simplicity, we show the dominant  one-loop  corrections to the  Higgs boson mass \cite{at}
\begin{eqnarray}
m_{h}^{2}  \simeq  M_{Z}^{2}\cos ^{2}2\beta \left( 1-\frac{3
}{8\pi ^{2}}\frac{m_{t}^{2}}{v^{2}}t\right)
+\frac{3}{4\pi ^{2}}\frac{m_{t}^{4}}{v^{2}}\left[ t+\frac{1}{2}X_{t}\right],
  \label{e1}
\end{eqnarray}%
where
\begin{eqnarray}
t =\log \left( \frac{M_{S}^{2}}{M_{t}^{2}}\right),  \,\,\,\, X_{t} =
\frac{2\widetilde{A}_{t}^{2}}{M_{S}^{2}}\left( 1-\frac{\widetilde{A}%
_{t}^{2}}{12M_{S}^{2}}\right),\,\,\,\, \widetilde{A}_{t}=A_{t}-\mu \cot \beta.
\end{eqnarray}%
Here  $A_t$ is
 the trilinear soft supersymmetry breaking (SSB) parameter associated with the top quark Yukawa coupling, $\mu$ is
the MSSM Higgs bilinear mixing term, $\cot \beta$ is the ratio of down and up-type Higgs vacuum expectation values (VEVs) and
$M_S=\sqrt{m^2_{Q_t}m^2_{U_t}}$ is the geometric mean of left and right
stop masses squared.
A 125 GeV Higgs mass  requires either a very large, ${\cal O} (\mathrm{few}-10)$ TeV,  stop quark mass \cite{Ajaib:2012vc}, or a large SSB trilinear $A_t$-term, with a stop quark mass of around a TeV \cite{Djouadi:2005gj}.

In the MSSM, through minimizing the tree level scalar potential,
the $Z$ boson mass $M_Z=91.2$ GeV   can be computed in terms of $\mu$ and the soft supersymmetry breaking mass terms for the up ($m_{H_{u}}$) and down ($m_{H_{d}}$)-type Higgs doublets \cite{mssm}
\begin{equation}
\frac{1}{2}M_{Z}^{2}= -\mu
^{2}+\left(\frac{m_{H_{d}}^{2} - m_{H_{u}}^{2}  \text{
tan}^{2}\beta}{\text{ tan}^{2}\beta -1} \right)\simeq -\mu
^{2}-m_{H_{u}}^{2}. \label{e4}
\end{equation}
The approximation in Eq. (\ref{e4}) works well for moderate and large $\tan\beta$ values.
We see from Eq. (\ref{e4}) that unless $\mu$ and $m_{H_{u}}$  values   are of order $M_Z$  some fine-tuning of these two parameters is required. As mentioned above, a 125 GeV Higgs in SUSY requires the stop quark mass to be around a TeV. Since  $H_u$ couples to the top quark, the heavy stop quark significantly affects the values of $m_{H_{u}}$ and pushes it to be order TeV at either the low or GUT scale,or at both scales simultaneously. We will show below that the heavy stop quark contribution to  $m_{H_{u}}$ can be canceled by a suitable choice of gaugino masses at the GUT scale and, in this way, ameliorate the little hierarchy problem.

Before discussing the effects of specific choices of non-universal gaugino masses
on the little hierarchy problem, we present a more general analysis with  arbitrary gaugino
masses   at the GUT scale. For this purpose we use a semi-analytic calculation for the MSSM
sparticle spectra presented in ref. \cite{Gogoladze:2009bd}.
\begin{eqnarray}
m_{H_{u}}^{2}
& \approx & -2.67 M_{3}^{2}+0.2 M_{2}^{2}-0.091 m_{0}^{2}-0.1 A_{t_0}^{2} -0.22 M_{3}M_{2} + \ldots, \label{h1}
\\
m_{Q_{t}}^{2}
& \approx &5.41 M_{3}^{2}+0.392 M_{2}^{2}+0.64 m_{0}^{2}
+0.115 M_{3}A_{t_{0}}
-0.072 M_{3}M_{2}+ \ldots, \label{h3}
\\
m_{U_{t}}^{2}
& \approx &4.52 M_{3}^2-0.188 M_{2}^2+0.273 m_{0}^2-0.066 A_{t_0}^2  -0.145 M_{3}M_{2} + \ldots,
 \label{h4}
\\
 A_{t} & \approx &-2.012 M_{3}-0.252 M_{2}+0.273 A_{t_0} + \ldots.
 \label{h2}
 \end{eqnarray}

The following GUT scale boundary conditions are employed

\begin{equation}
\alpha _{G} = {1}/{24.32},\,\, \,\, M_{G}=2.0\times 10^{16} {\rm GeV},\,\,\,\, y_{t}(M_{G})=0.512
\end{equation}
  Here $\alpha_2 = \alpha_1 = \alpha_G$ and we do not enforce exact unification $\alpha_3 = \alpha_2 =
\alpha_1$ at the GUT scale, since a few percent deviation from the
unification condition can be expected due to unknown GUT scale threshold
corrections \cite{Hisano:1992jj}. $M_{G}$ and $y_{t}$ stand for GUT scale and top Yukawa coupling.
By integrating the one loop RGEs \cite{RGE},  the MSSM sparticle masses at  $M_{Z}$ scale can be expressed   in terms
of the GUT scale fundamental parameters $m_0, M_{1,2,3},
A_{t_{0}}$ and the $\mu$ term. Here $m_0$ stands for  universal SSB mass term for sfermions, $M_{1,2,3}$ are SSB gaugino masses  and $A_{t_{0}}$ is the trilinear SSB coupling.

The ellipses denote additional terms which do not give significant contribution to the given sparticle masses. In the results presented later, which are based on the numerical calculation employed by ISAJET 7.84 \cite{ISAJET}, full two loop RGE's are used to calculate the parameters given in Eqs. (\ref{h1})-(\ref{h2}).

We observe from Eq.(\ref{h1}) that the simplest way  to  reduce the absolute
value of $m_{H_{u}}^{2}$ (which, for our purpose,
can be regarded as a measure of the fine-tuning)  is  to have comparable values
for the first two terms. This suggests the need for non-universal
gaugino masses at $M_{G}$, with $\vert M_2 \vert > \vert M_3 \vert$.
At the same time,  from Eqs. (\ref{h3}) and (\ref{h4}) we see that it is possible to have large values of the stop quark mass.

It is interesting to note that the relative signs of  $M_3$ and $M_2$
play an important role.
It follows from Eq.(\ref{h2}) that opposite signs for $M_2$ and
$M_3$  reduce the absolute value of $A_t$, after RGE running from high to low scale. This, however,
reduces  $m_h$ and to compensate for this reduction  one should increase the
value of $M_3$ at $M_{\rm GUT}$. This, in turn, increases the absolute value
of $m_{H_{u}}^{2}$ at  $M_{SUSY}$.
Thus,
from Eq. (\ref{h1}) we find that in order to reduce the amount of fine-tuning,
the following  condition should be met,
\begin{eqnarray}
 M_2/M_3  \approx \sqrt{2.67/0.2}\approx 3.6\,.  \label{rr1}
\end{eqnarray}
The results in Eq. (\ref{rr1}) are obtained using one loop RGE's, but as we will see later on, the ratio $ M_2/M_3  \approx 3.6$  is in good agreement  with our finding $ M_2/M_3  \approx 3.3$ using the ISAJET 7.84  package which employs two loop RGE's.

\section{Fine-Tuning Constraints \label{ft}}
The latest (7.84) version of  ISAJET \cite{ISAJET} calculates the  fine-tuning conditions related to the little hierarchy problem at $M_{EW}$
and at the GUT scale ($M_{HS}$). We will briefly describe these parameters in this section.

After including the one-loop effective potential contributions to the tree level MSSM Higgs potential, $M_Z$ the Z boson mass is given by the following relation:
\begin{equation}
\frac{M_Z^2}{2} =
\frac{(m_{H_d}^2+\Sigma_d^d)-(m_{H_u}^2+\Sigma_u^u)\tan^2\beta}{\tan^2\beta
-1} -\mu^2 \; .
\label{eq:mssmmu}
\end{equation}
The $\Sigma$'s stand for the contributions coming from the one-loop effective potential (For more details see ref. \cite{Baer:2012mv}). All parameters  in Eq. (\ref{eq:mssmmu}) are defined at the weak scale $M_{EW}$.

\subsection{Electroweak Scale Fine-Tuning}
\label{esft}

In order to measure the EW scale fine-tuning condition associated with the little hierarchy problem, the following definitions are used \cite{Baer:2012mv}:
\begin{equation}
 C_{H_d}\equiv |m_{H_d}^2/(\tan^2\beta -1)|,\,\, C_{H_u}\equiv
|-m_{H_u}^2\tan^2\beta /(\tan^2\beta -1)|, \, \, C_\mu\equiv |-\mu^2 |,
\label{cc1}
\end{equation}
 with
each $C_{\Sigma_{u,d}^{u,d} (i)}$  less
than some characteristic value of order $M_Z^2$.
Here, $i$ labels the SM and supersymmetric
particles that contribute to the one-loop Higgs potential.
For the fine-tuning condition we have
\begin{equation}
 \Delta_{\rm EW}\equiv {\rm max}(C_i )/(M_Z^2/2).
\label{eq:ewft}
\end{equation}
Note that Eq. (\ref{eq:ewft}) defines the fine-tuning  condition  at $M_{EW}$ without addressing
the question of the origin of the parameters that are involved.

\subsection{High Scale Fine-Tuning}
\label{hsft}

In most SUSY breaking scenarios the parameters in
Eq.~(\ref{eq:mssmmu}) are defined at a scale higher than $M_{EW}$.
In order to fully address  the fine-tuning condition we need to  check the relations
 among the parameters involved  in Eq.~(\ref{eq:mssmmu}) at high scale. We relate the parameters at low and high scales as follows:
 \begin{equation}
 m_{H_{u,d}}^2=
m_{H_{u,d}}^2(M_{HS}) +\delta m_{H_{u,d}}^2, \ \,\,\,
\mu^2=\mu^2(M_{HS})+\delta\mu^2.
\end{equation}
 Here
$m_{H_{u,d}}^2(M_{HS})$ and $\mu^2(M_{HS})$ are the corresponding
parameters renormalized at the high scale, and
$\delta m_{H_{u,d}}^2$, $\delta\mu^2$ measure how the given parameter is changed due to renormalization group evolution (RGE).
 Eq.~(\ref{eq:mssmmu}) can be re-expressed in the form
\begin{eqnarray}
\frac{m_Z^2}{2} &=& \frac{(m_{H_d}^2(M_{HS})+ \delta m_{H_d}^2 +
\Sigma_d^d)-
(m_{H_u}^2(M_{HS})+\delta m_{H_u}^2+\Sigma_u^u)\tan^2\beta}{\tan^2\beta -1}
\nonumber \\
&-& (\mu^2(M_{HS})+\delta\mu^2)\;.
\label{eq:FT}
\end{eqnarray}
Following ref. \cite{Baer:2012mv}, we  introduce the following parameters:
\begin{eqnarray}
&B_{H_d}\equiv|m_{H_d}^2(M_{HS})/(\tan^2\beta -1)|,
B_{\delta H_d}\equiv |\delta m_{H_d}^2/(\tan^2\beta -1)|, \nonumber \\
&B_{H_u}\equiv|-m_{H_u}^2(M_{HS})\tan^2\beta /(\tan^2\beta -1)|, B_{\mu}\equiv|\mu^2(M_{HS})|, \nonumber \\
&B_{\delta H_u}\equiv|-\delta m_{H_u}^2\tan^2\beta /(\tan^2\beta -1)|,
  B_{\delta \mu}\equiv |\delta \mu^2|,
  \label{bb1}
\end{eqnarray}
and  the
high scale fine-tuning measure $\Delta_{\rm HS}$ is defined to be
\begin{equation}
\Delta_{\rm HS}\equiv {\rm max}(B_i )/(M_Z^2/2).
\label{eq:hsft}
\end{equation}

The current experimental bound on the chargino mass ($m_{\widetilde W}> 103$ GeV) \cite{Nakamura:2010zzi} indicates that either $\Delta_{EW}$ or $\Delta_{HS}$ cannot be less than 1. The quantities $\Delta_{EW}$ and  $\Delta_{HS}$ measure the sensitivity of the Z-boson mass to the parameters defined in Eqs. (\ref{cc1}) and (\ref{bb1}), such that $(100/\Delta_{EW})\%$  ($(100/\Delta_{HS})\%$) is the degree of fine-tuning at the corresponding scale.


\section{ The Model \label{model}}

In  Section \ref{results} we will show that the little hierarchy problem is
largely resolved if the MSSM is embedded in the 4-2-2 model \cite{Pati:1974yy}. It
seems natural to assume that in 4-2-2  the asymptotic (high/GUT  scale) gaugino masses
associated with $SU(4)_c, SU(2)_L$, and  $SU(2)_R$ are three
independent parameters. This can be reduced to two independent parameters
in the presence of C  parity \cite{c-parity}, and we can have the
following asymptotic relation among the MSSM
gaugino masses:
\begin{equation}
M_1 = {\frac{2}{5}}M_3+{\frac{3}{5}}M_2, \label{PSMR}
\end{equation}
where $M_3, M_2, M_1$ denote the $SU(3)_c, SU(2)_L$ and $U(1)_Y$
asymptotic gaugino masses respectively. Reducing the parameters this way by keeping $M_2$ and $M_3$ independent while calculating  $M_1$ using Eq. (\ref{PSMR})   in the gaugino sector is also phenomenologically
motivated. The reason for this is that  except for the bino, all remaining sparticle masses weakly depend on the value of $M_1$, especially if the neutralino is the lightest supersymmetric particle (LSP).
 We will
assume that due to C-parity the SSB
mass terms induced at the high scale through gravity mediated supersymmetry breaking \cite{Chamseddine:1982jx} are equal in magnitude for the  squarks and sleptons  within the family. For simplicity and for comparison with the CMSSM case, we also assume
$m_0=m_{H_{u}}=m_{H_{d}}$.
Thus, we have the following fundamental parameters in the SSB sector:
\begin{align}
m_{0}, m_{H_u}, m_{H_d}, M_2, M_3, A_0, \tan\beta.
\label{params}
\end{align}
 In this paper we mainly focus on $\mu>0$.
 Although not required, we will
assume that the gauge coupling unification condition $g_3=g_1=g_2$
holds at $M_{\rm GUT}$ in 4-2-2. Such a scenario can arise,
for example, from a higher dimensional $SO(10)$
\cite{Hebecker:2001jb} or $SU(8)$ \cite{su8} model after suitable
compactification.


\section{Scanning Procedure and Phenomenological Constraints \label{constraintsSection}}

We employ the ISAJET~7.84 package~\cite{ISAJET}  to perform random
scans over the fundamental parameter space. In this package, the weak scale values of gauge and third generation Yukawa
couplings are evolved to the GUT scale via the MSSM renormalization
group equations (RGEs) in the $\overline{DR}$ regularization scheme.
We do not strictly enforce the unification condition  at $M_{\rm
GUT}$, since a few percent deviation from unification can be
assigned to unknown GUT-scale threshold
corrections~\cite{Hisano:1992jj}.
The deviation between $g_1=g_2$ and $g_3$ at $M_{\rm GUT}$ is no
worse than $3-4\%$.
For simplicity,  we do not include the Dirac neutrino Yukawa coupling
in the RGEs. For possible values of this coupling, the impact on our calculations is not significant.

The various boundary conditions are imposed at the
GUT scale and all the SSB
parameters, along with the gauge and Yukawa couplings, are evolved
back to the weak scale $M_{\rm Z}$.
In the evolution of Yukawa couplings the SUSY threshold
corrections~\cite{Pierce:1996zz} are taken into account at the
common scale $M_{\rm SUSY}= \sqrt{m_{{\tilde t}_L}m_{{\tilde t}_R}}$,
where $m_{{\tilde t}_L}$ and $m_{{\tilde t}_R}$
denote the masses of the third generation left and right-handed stop quarks.
The entire parameter set is iteratively run between $M_{\rm Z}$ and $M_{\rm
GUT}$ using the full 2-loop RGEs until a stable solution is
obtained. To better account for leading-log corrections, one-loop
step-beta functions are adopted for the gauge and Yukawa couplings, and
the SSB parameters $m_i$ are extracted from RGEs at multiple scales
$m_i=m_i(m_i)$. The RGE-improved 1-loop effective potential is
minimized at $M_{\rm SUSY}$, which effectively
accounts for the leading 2-loop corrections. Full 1-loop radiative
corrections are incorporated for all sparticle masses.

We have performed random scans for the following range of the parameter space:
\begin{align}
0\leq  m_{0}=m_{H_{u}}=m_{H_{d}}  \leq 20\, \rm{TeV} \nonumber \\
0 \leq M_{3}  \leq 5 \, \rm{TeV} \nonumber \\
0 \leq M_{2}  \leq 5 \, \rm{TeV}  \nonumber \\
2\leq \tan\beta \leq 60 \nonumber \\
-3\leq A_{0}/m_{0} \leq 3\nonumber\\
\mu > 0
 \label{parameterRange}
\end{align}
We set    $m_t = 173.3\, {\rm GeV}$  \cite{Aaltonen:2012ra}
and  $m_b(m_Z)=2.83$ GeV, which is hard-coded into ISAJET.

In performing the random scan a uniform and logarithmic distribution of random points is first generated in the parameter space given in Eq. (\ref{parameterRange}).
The function RNORMX \cite{Leva} is then employed
to generate a gaussian distribution around each point in the parameter space.
The collected data points all satisfy
the requirement of REWSB with one of the neutralino being the lightest supersymmetric particle (LSP).

After collecting the data, we impose
the mass bounds on all the particles \cite{Nakamura:2010zzi} and use the
IsaTools package~\cite{Baer:2002fv}
to implement the various phenomenological constraints. We successively apply the following experimental constraints on the data that
we acquire from ISAJET:
\begin{table}[h]\centering
\begin{tabular}{rlc}
$1.7 \times 10^{-9} \leq\, BR(B_s \rightarrow \mu^+ \mu^-) $&$ \leq\, 4.7 \times 10^{-9}$        & \cite{Aaij:2012hcp}     \\
$2.85 \times 10^{-4} \leq BR(b \rightarrow s \gamma) $&$ \leq\, 4.24 \times 10^{-4} \;
 (2\sigma)$ &   \cite{Barberio:2008fa}  \\
$0.15 \leq \frac{BR(B_u\rightarrow
\tau \nu_{\tau})_{\rm MSSM}}{BR(B_u\rightarrow \tau \nu_{\tau})_{\rm SM}}$&$ \leq\, 2.41 \;
(3\sigma)$ &   \cite{Barberio:2008fa}  \\
 $ 0 \leq \Delta(g-2)_{\mu}/2 $ & $ \leq 55.6 \times 10^{-10} $ & \cite{Bennett:2006fi}
\end{tabular}\label{table}
\end{table}

Note that for $\Delta(g-2)_{\mu}$, we only require that the  model does no worse than the SM.


\begin{figure}[th!]
\begin{center}
\includegraphics[width=7.6cm,height=6.cm]{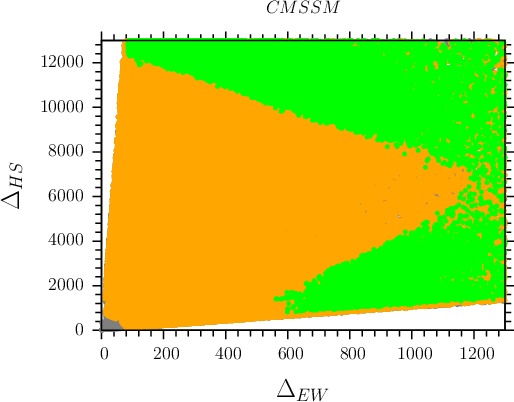}
\includegraphics[width=7.6cm,height=6.cm]{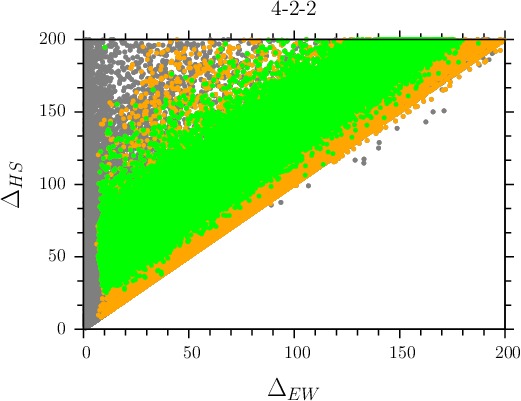}
\end{center}
\caption{Plots in  the $\Delta_{HS}- \Delta_{EW}$ planes for CMSSM and 4-2-2 cases. Gray points are consistent with REWSB and neutralino to be LSP. The orange points form a subset of the
gray ones and satisfy all the constraints described in Section \ref{constraintsSection}.  Green points belong to the subset of orange points
and satisfy the Higgs mass range $123\, {\rm GeV} \leq m_h \leq 127 \,{\rm GeV}$.}
\label{fig1}
\end{figure}


\begin{figure}[t]
\begin{center}
\includegraphics[width=7.2cm,height=6.cm]{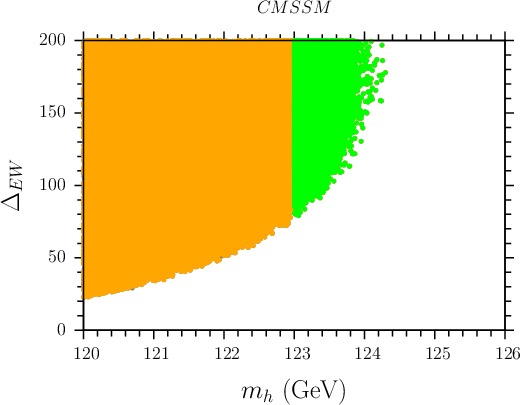}
\includegraphics[width=7.2cm,height=6.cm]{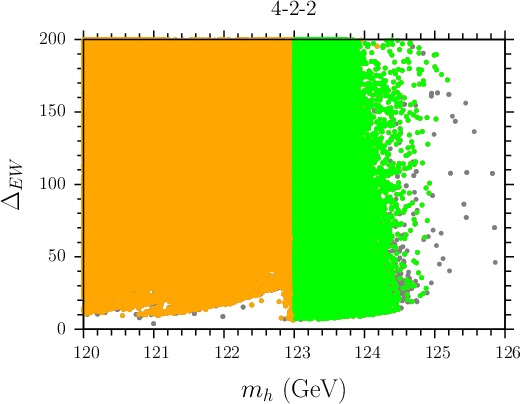}\vspace*{3mm}
\includegraphics[width=7.2cm,height=5.6cm]{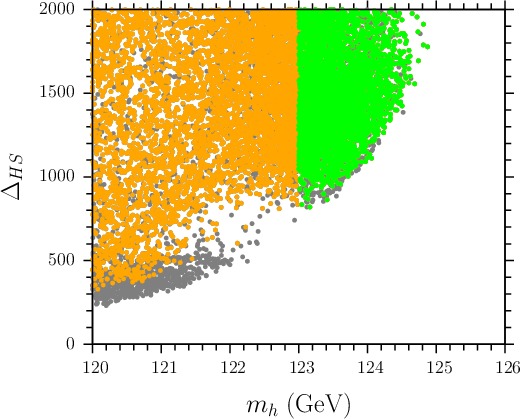}
\includegraphics[width=7.2cm,height=5.6cm]{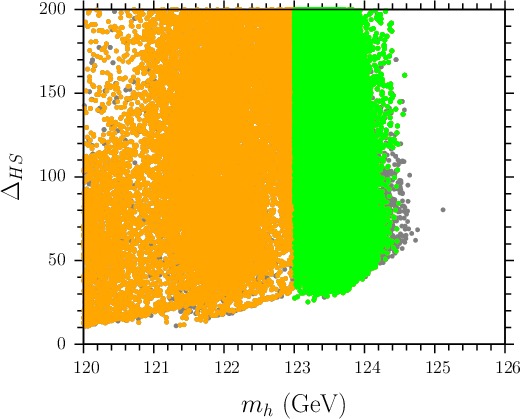}
\end{center}
\caption{Plots in $\Delta_{EW} - m_{h}$ and $\Delta_{HS} - m_{h}$ planes for CMSSM and 4-2-2. Color coding
is the same as described in Figure \ref{fig1}.  }
\label{fig2}
\end{figure}


\begin{figure}[t!]
\begin{center}
\includegraphics[width=7.2cm,height=6.cm]{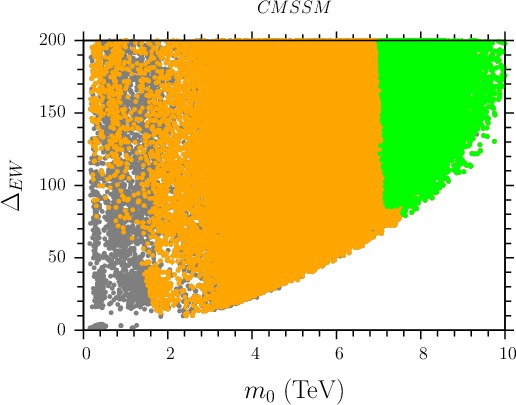}
\includegraphics[width=7.2cm,height=6.cm]{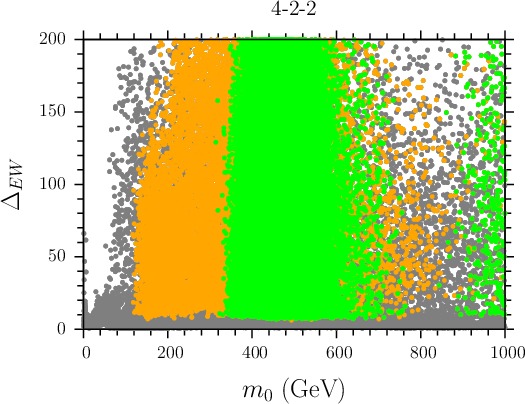}\vspace*{3mm}
\includegraphics[width=7.2cm,height=5.6cm]{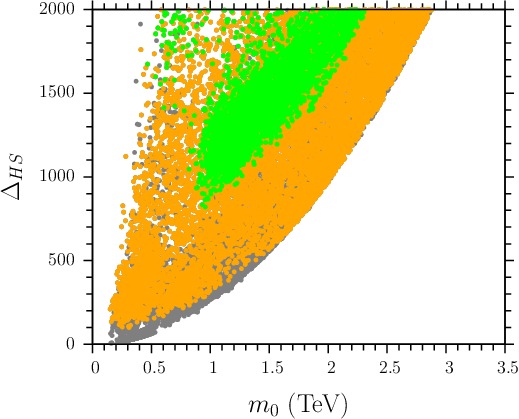}
\includegraphics[width=7.2cm,height=5.6cm]{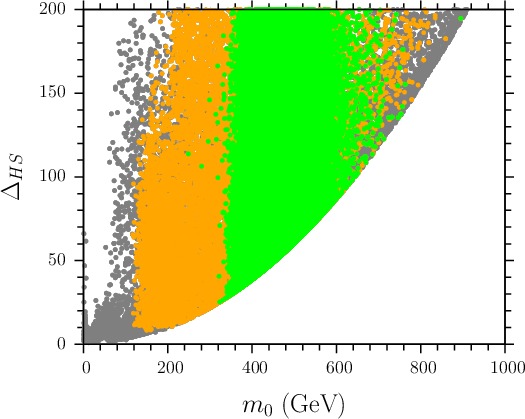}
\end{center}
\caption{Plots in $\Delta_{EW} - m_{0}$ and $\Delta_{HS} - m_{0}$ planes for CMSSM and 4-2-2 cases. Color coding
is the same as described in Figure \ref{fig1}.   }
\label{fig3}
\end{figure}


\begin{figure}[t!]
\begin{center}
\includegraphics[width=7.2cm,height=6.cm]{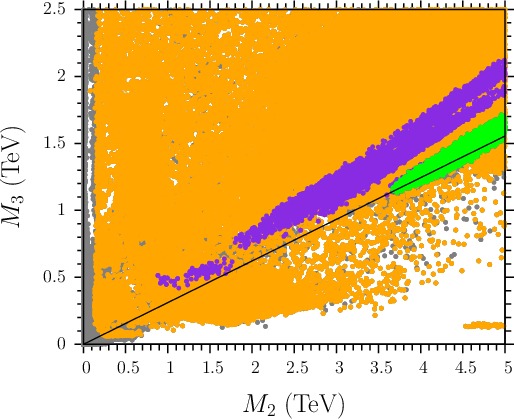}
\includegraphics[width=7.2cm,height=6.cm]{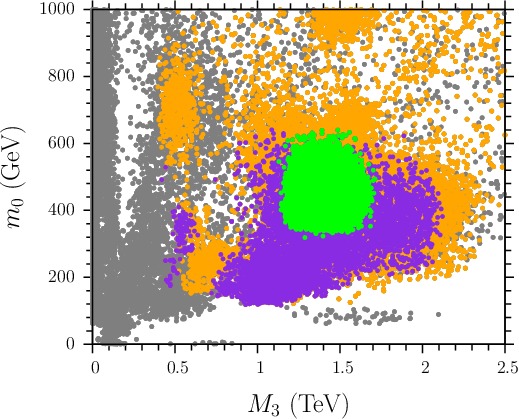}
\end{center}
\caption{Plots in  the $M_3- M_2$  and $m_0-M_3$ planes for 4-2-2. Gray points are consistent with REWSB and neutralino LSP. The orange points, a subset of the
gray ones, satisfy all the constraints described in Section \ref{constraintsSection}.
 Purple points are subset of orange points with $\Delta_{EW}\leq 100$ and $\Delta_{HS}\leq 100$.
 Green points belong to a subset of purple points
with the Higgs mass range $123\, {\rm GeV} \leq m_h \leq 127 \,{\rm GeV}$. }
\label{fig4}
\end{figure}


\begin{figure}[t!]
\begin{center}
\includegraphics[width=7.2cm,height=6.cm]{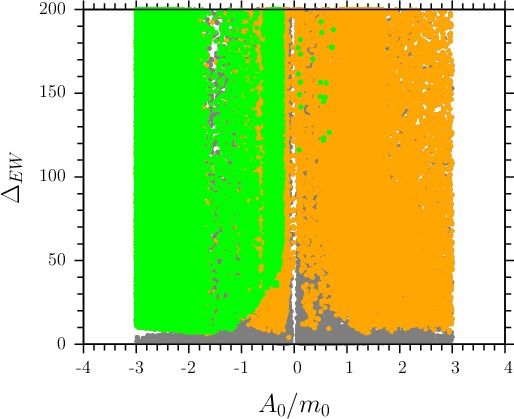}
\includegraphics[width=7.2cm,height=6.cm]{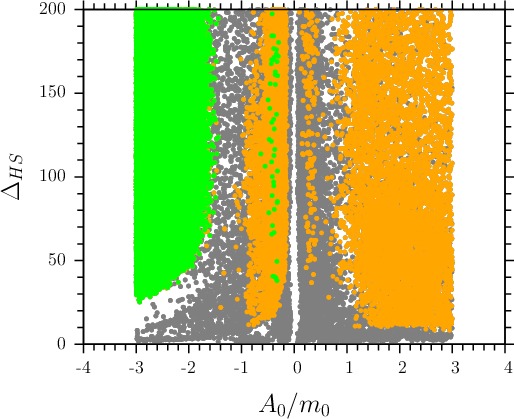}\vspace*{3mm}
\includegraphics[width=7.2cm,height=5.6cm]{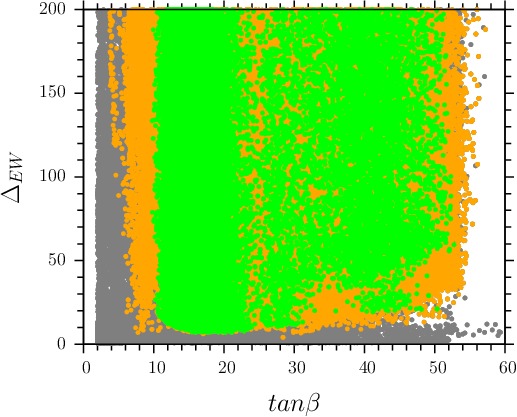}
\includegraphics[width=7.2cm,height=5.6cm]{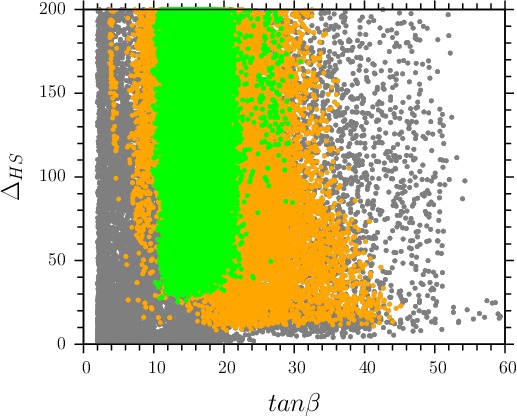}
\end{center}
\caption{Plots in $\Delta_{EW} - A_0/m_{0}$,  $\Delta_{HS} - A_0/m_{0}$, $\Delta_{EW} - \tan\beta$,  $\Delta_{HS} - \tan\beta$ planes for 4-2-2. Color coding
is the same as described in Figure \ref{fig1}.}
\label{fig5}
\end{figure}


\begin{figure}[t!]
\begin{center}
\includegraphics[width=7.2cm,height=6.cm]{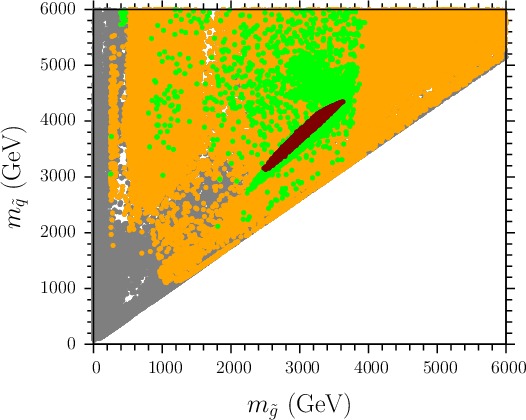}
\includegraphics[width=7.2cm,height=6cm]{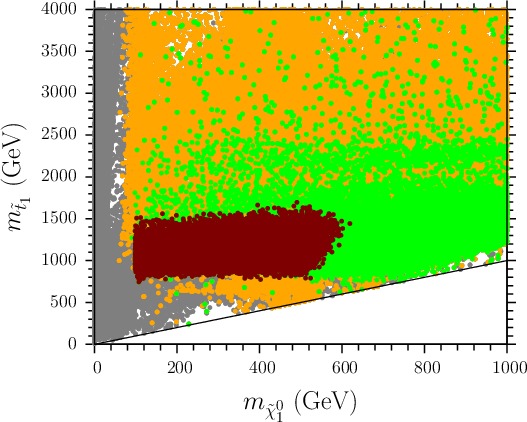}\vspace*{3mm}
\includegraphics[width=7.2cm,height=6cm]{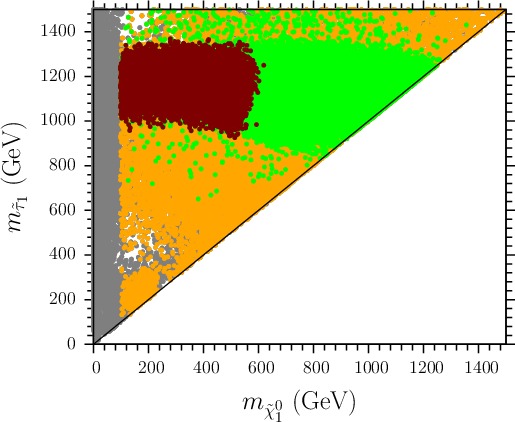}
\includegraphics[width=7.2cm,height=6.cm]{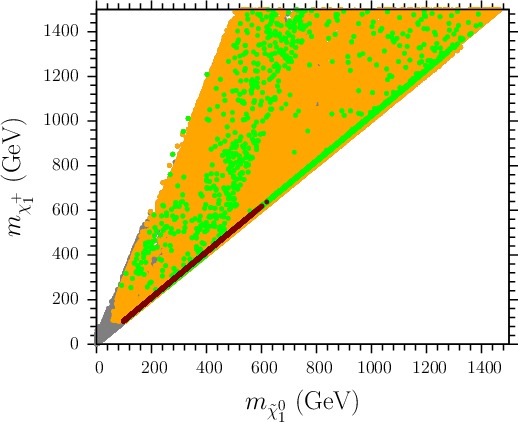}
\end{center}
\caption{Plots in  the
$m_{\tilde{q}}-m_{\tilde{g}}$, $m_{\tilde{\chi}^{+}_1}-m_{\tilde{\chi}^{0}_1}$, $m_{\tilde{t}_{1}}-m_{\tilde{\chi}^{0}_1}$ and $m_{\tilde{\tau}_{1}}-m_{\tilde{\chi}^{0}_1}$ planes for 4-2-2. Color coding
is the same as described in Figure \ref{fig1}. In addition, we have used maroon color to denote a subset of the green points with $\Delta_{HS}<100$ and $\Delta_{EW}<100$. }
\label{fig6}
\end{figure}


\begin{figure}[t!]
\begin{center}
\includegraphics[width=7.2cm,height=6.cm]{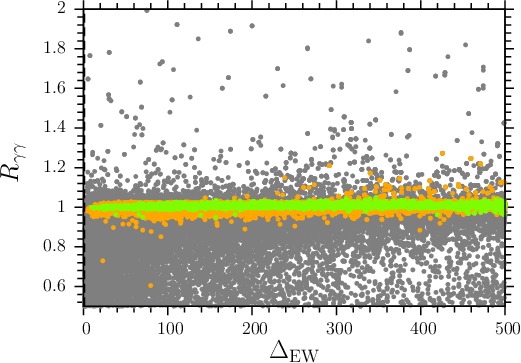}
\includegraphics[width=7.2cm,height=6.cm]{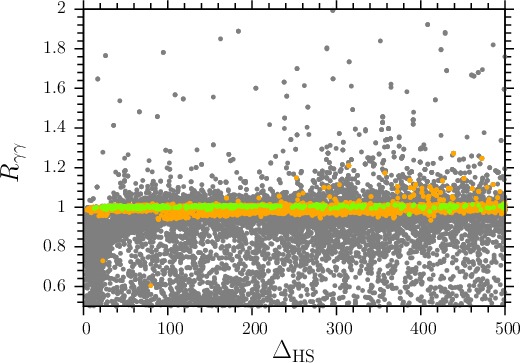}
\end{center}
\caption{Plots in $R_{\gamma \gamma} - \Delta_{EW}$,  and $R_{\gamma \gamma} - \Delta_{HS}$ planes for 4-2-2 case. Color coding
is the same as described in Figure \ref{fig1}.  }
\label{fig7}
\end{figure}


\section{Results \label{results}}


In Figures \ref{fig1}, \ref{fig2} and \ref{fig3} we compare the CMSSM  with the 4-2-2 model .  Figure \ref{fig1}  shows our results in the $\Delta_{EW} - \Delta_{HS}$ planes for the CMSSM and 4-2-2 models.
The gray points are consistent with REWSB and LSP neutralino. The orange points, a subset of the
gray ones, satisfy all the constraints described in Section \ref{constraintsSection}. The green points correspond
to the light CP-even  Higgs mass range
\begin{align}
123\, {\rm GeV} \leq m_h \leq 127 \,{\rm GeV},
\label{hm}
\end{align}
and form a subset of the orange points.
We assume  that the Higgs boson mass range in Eq. (\ref{hm}) is compatible with the current experimental bound, while allowing  for a 2-3 GeV error margin in the RG-improved one-loop effective potential calculation \cite{Degrassi:2002fi}.
Note that  varying the top quark mass within a $1\,\sigma$ interval can increase  the Higgs boson mass by 1 GeV or so \cite{Gogoladze:2012ii}.

  We do not apply the WMAP bound \cite{Komatsu:2008hk} on the relic LSP abundance since the neutralino is mostly a higgsino-like particle for the natural SUSY case. This usually yields small values for the thermal LSP relic abundance, unless the LSP neutralino mass is around 1 TeV or so \cite{Baer:2006te}.  In ref. \cite{Baer:2012uy}
 the axion mechanism  has been invoked to yield the desired relic abundance for axion-nonthermal LSP dark matter.

The green points in the  $\Delta_{EW} - \Delta_{HS}$ plane for CMSSM indicate that it is possible to have $\Delta_{EW}\approx 50$ (green points upper left corner), but with $\Delta_{HS}> 12, 000\, (0.008\%)$. On the other hand in the CMSSM for minimal value of  $\Delta_{HS}$, namely $\Delta_{HS}\approx 800\, (0.12\%)$ (fine-tuning), we also  find a minimal value for $\Delta_{EW}\approx 600\, (0.17\%)$.  This observation shows how important it is to simultaneously study    both the low and high scale  fine-tuning conditions for the little hierarchy problem in order to have the complete picture.

We now focus on the 4-2-2 model, which does much better than the CMSSM in terms of the little hierarchy problem.
We find that for 4-2-2 it is possible to have solutions with  much smaller values for $\Delta_{EW}$ and $\Delta_{HS}$  that are consistent with  all collider constraints described in Section
 \ref{constraintsSection}, and  with the light CP even Higgs having mass in the interval presented in Eq. (\ref{hm}). Thus, $\Delta_{EW}$  can be as small as 10, with $\Delta_{HS}\approx 20$.

 In Figure \ref{fig2} we present the results in $\Delta_{EW}-m_h$ and $\Delta_{HS} - m_h$ planes for the CMSSM and 4-2-2
 cases. The color coding is the same as in Figure {\ref{fig1}}. Green points in the $\Delta_{EW}-m_h$ plane can be as low as 80  for CMSSM, but as we learned from Figure \ref{fig1}, these points correspond to  $\Delta_{HS}>12, 000$. Note that the fine-tuning parameters $\Delta_{EW}$ and $\Delta_{HS}$ in CMSSM rapidly change with changes in the light CP even Higgs boson mass, especially when $m_h>123$ GeV.  In contrast, the 4-2-2  model shows a very mild dependence of  $\Delta_{EW}$ and $\Delta_{HS}$  on the Higgs boson mass up to the 125 GeV.

In Figure \ref{fig3} we show how the fine-tuning parameters $\Delta_{EW}$ and $\Delta_{HS}$ differ in their dependence on the SSB mass parameter $m_0$. We present  both the CMSSM and 4-2-2 cases in this figure.
 For CMSSM, the $\Delta_{EW} \ vs. \ m_0$ plane shows that minimal values for $\Delta_{EW}$ correspond to $m_0\approx 7$ TeV. On the other hand,  we see from the $\Delta_{HS}-m_0$ figure that the minimal values for $\Delta_{HS}$  are achieved for
 $m_0\approx 1 \tev$.

 We can observe a different relationship among these parameters in the  $\Delta_{EW}-m_0$ and $\Delta_{HS}-m_0$ planes in Figure \ref{fig3} for the 4-2-2 case.
In the $\Delta_{EW}-m_0$ plane we see that $\Delta_{EW}$  depends  very mildly on the values of $m_0$, for $0<m_0<1$ TeV. The $\Delta_{HS}-m_0$ plane shows that   $\Delta_{HS}$ varies significantly as $m_0$ increases. We also observe from Figure \ref{fig3} that in the 4-2-2 case, $m_0$ can be as light as 400 GeV with
$m_h>123$ GeV and relatively small $\Delta_{EW}$ and $\Delta_{HS}$ values. This finding can be understood from Figure \ref{fig4} where results are presented in $M_3-M_2$ and $m_{0}- M_{3}$ planes.

In Figure \ref{fig4} the gray points are consistent with REWSB and LSP neutralino. The orange points form a subset of the
gray ones and satisfy all the constraints described in Section \ref{constraintsSection}.
THe purple points form a subset of the orange points with $\Delta_{EW}\leq 100$ and $\Delta_{HS}\leq 100$ conditions.
The green points belong to a subset of purple points
and satisfy the Higgs mass range $123\, {\rm GeV} \leq m_h \leq 127 \,{\rm GeV}$.

From the $M_3-M_2$ plane we can see that some green points lie near the line corresponding to $M_{2}/M_{3} \approx 3.3$.
As  discussed in Section \ref{lhp}, in order to have a cancelation between $M_2$ and $M_3$ in the expression for  $M_{H_{u}}$,
 the ratio needs to satisfy $M_2/M_3\approx 3.6$. This result was obtained  using one-loop RGE's for the SSB parameters (see Eqs. (\ref{h1})-(\ref{h4})), but still serves as a good approximation to describe the  green points in $M_3-M_2$ plane.
At the same time we can see from Eqs. (\ref{h3}) and (\ref{h4}) that having $M_2/M_3\approx 3.6$ does not cancel the $M_3$ and $M_2$ contributions in the expressions for the stop quark, and so the stop quark mass can still be large. As we know, a large stop quark mass is required to provide significant radiative contributions that yield a light CP-even Higgs mass of around 125 GeV.

We show that the interval
$400$ GeV $< m_0<$ 600 GeV and $1.1$ TeV $<M_3<$ $1.7$ TeV in 4-2-2 model is in good agreement with the current experimental bounds.

In Figure \ref{fig5} we show how the parameter space changes if solution of the little hierarchy problem
 is required  at the low or  high  scale. We present the results in $\Delta_{EW}-A_0/m_0$, $\Delta_{HS}-A_0/m_0$, $\Delta_{EW}-\tan\beta$ and $\Delta_{HS}-\tan\beta$ planes. Color coding is  same as in Figure \ref{fig1}.
We see from the $\Delta_{EW}-A_0/m_0$ plane that requiring only low scale fine-tuning ($\Delta_{EW}<100$) restricts the sign of the ratio $A_0/m_0$.  In other words we can have a solution to the little hierarchy problem for any negative value of $A_0$. For the high scale fine-tuning case ($\Delta_{HS}<100$),  we have a relatively narrow interval $-3<A_{0}/m_{0}<-1.5$.
A similar situation occurs for the $\tan\beta$ case, where the  allowed  interval for  $\tan\beta$ changes significantly when we compare low scale fine-tuning with the high scale one. The $\Delta_{EW}-\tan\beta$ plane indicates that there is no little hierarchy problem for $10<\tan\beta< 55$.  On the other hand,  from  the $\Delta_{HS}-\tan\beta$ plane we see that the range $10< \tan\beta<22$ has $\Delta_{HS}< 100$ and $123\, {\rm GeV} \leq m_h \leq 127 \,{\rm GeV}$.

Note that this also shows that precise   $t-b-\tau$ Yukawa unification is not   compatible with the 4-2-2 model if we insist on the resolution of the little hierarchy problem. This is because, the solution of the little hierarchy  problem prefers $\tan\beta<20$,  while precise $t-b-\tau$ Yukawa unification predicts $\tan\beta\approx 50$ \cite{Gogoladze:2011aa}.

\begin{table}[t!]\vspace{0.1cm}
\centering
\begin{tabular}{|p{3cm}|p{3cm}p{3cm}p{3cm}|}
\hline
\hline

                 	&	Point 1	&	Point 2	&	Point 3	\\
\hline

$m_0$	&$	          1086.39	$&$	          460.72	$&$	          497.64	$\\
$M_1$	&$	          979.1	$&$	          3313.58	$&$	          3606.12	$\\
$M_2$	&$	          979.1	$&$	          4579.38	$&$	          4908.89	$\\
$M_3$	&$	          979.1	$&$	          1414.89	$&$	          1651.94	$\\
$A_0$	&$	         -3244.79	$&$	         -1270.88	$&$	         -1390.14	$\\
$\tan\beta$      	&$	28.49	$&$	15.41	$&$	16.47	$\\
\hline		  		  		  	
$\mu$            	&$	1853	$&$	176	$&$	746	$\\

\hline		  		  		  	
$m_h$            	&$	124.06	$&$	124	$&$	124.1	$\\
$m_H$            	&$	1862	$&$	2856	$&$	3109	$\\
$m_A$            	&$	1850	$&$	2838	$&$	3088	$\\
$m_{H^{\pm}}$    	&$	1864	$&$	2857	$&$	3110	$\\
		  		  		  	
\hline		  		  		  	
$m_{\tilde{\chi}^0_{1,2}}$	&$	         424,          807	$&$	         180,          182	$&$	         759,          762	$\\

$m_{\tilde{\chi}^0_{3,4}}$	&$	        1845,         1847	$&$	        1477,         3757	$&$	        1620,         4032	$\\

$m_{\tilde{\chi}^{\pm}_{1,2}}$	&$	         810,         1850	$&$	         188,         3754	$&$	         780,         4023	$\\

$m_{\tilde{g}}$  	&$	2180	$&$	3048	$&$	3515	$\\
		  		  		  	
\hline $m_{ \tilde{u}_{L,R}}$	&$	        2239,         2174	$&$	        3842,         2719	$&$	        4253,         3118	$\\
                 		  		  		  	
$m_{\tilde{t}_{1,2}}$	&$	        1084,         1744	$&$	        1039,         3394	$&$	        1467,         3768	$\\
                 		  		  		  	
\hline $m_{ \tilde{d}_{L,R}}$	&$	        2240,         2166	$&$	        3843,         2629	$&$	        4254,         3025	$\\
                 		  		  		  	
$m_{ \tilde{b}_{1,2}}$	&$	        1721,         1947	$&$	        2524,         3436	$&$	        2905,         3808	$\\
                 		  		  		  	
\hline		  		  		  	
$m_{\tilde{\nu}_{1}}$	&$	1261	$&$	2980	$&$	3182	$\\
                 		  		  		  	
$m_{\tilde{\nu}_{3}}$	&$	1098	$&$	2972	$&$	3164	$\\
                 		  		  		  	
\hline		  		  		  	
$m_{ \tilde{e}_{L,R}}$	&$	        1265,         1144	$&$	        2978,         1296	$&$	        3181,         1407	$\\
                		  		  		  	
$m_{\tilde{\tau}_{1,2}}$	&$	         719,         1107	$&$	        1189,         2961	$&$	        1276,         3156	$\\
                		  		  		  	
\hline		  		  		  	
		  		  		  	
$\sigma_{SI}({\rm pb})$	&$	  9.24\times 10^{-12}	$&$	  1.79\times 10^{-10}	$&$	  2.84\times 10^{-10}	$\\

$\sigma_{SD}({\rm pb})$	&$	  2.46\times 10^{-09}	$&$	  2.29\times 10^{-06}	$&$	  2.36\times 10^{-07}	$\\

$\Omega_{CDM}h^{2}$	&$	  7.06	$&$	  0.007	$&$	  0.11	$\\
\hline							
$\Delta_{EW}$	&$	  827	$&$	  15.4	$&$	  134	$\\
		  		  		  	
$\Delta_{HS}$	&$	  1110	$&$	  51.3	$&$	  181	$\\

\hline
\hline
\end{tabular}
\caption{ Point 1 displays solution with  minimal value of $\Delta_{EW}$
and $\Delta_{HS}$  in the framework of CMSSM. Point 2 represents minimal value of $\Delta_{EW}$
and $\Delta_{HS}$ in the 4-2-2 model. Point 3 depict solutions corresponding minimal $\Delta_{EW}$ and
$\Delta_{HS}$ and best $\Omega_{CDM}h^{2}$ values. All benchmark points satisfying the various constraints
mentioned in Section \ref{constraintsSection}
}
\label{tab1}
\end{table}

In Figure \ref{fig6}, we present the SUSY particle spectrum  in the $m_{\tilde{q}}-m_{\tilde{g}}$,  $m_{\tilde{t}_{1}}-m_{\tilde{\chi}^{0}_1}$, $m_{\tilde{\tau}_{1}}-m_{\tilde{\chi}^{0}_1}$
and $m_{\tilde{\chi}^{+}_1}-m_{\tilde{\chi}^{0}_1}$ planes.
Color coding
is the same as described in Figure \ref{fig1}. In addition, we have used maroon color to denote a subset of the green points, that have $\Delta_{HS}<100$ and $\Delta_{EW}<100$.  The $m_{\tilde{q}}-m_{\tilde{g}}$ panel shows that in the 4-2-2 model,  resolution of the little hierarchy problem with acceptable Higgs boson  mass  yields a  gluino mass in the interval $2800$ GeV $<m_{\tilde{g}}<$ $4000$ GeV and the first two family squark masses in the  range
3000 GeV $<m_{\tilde{q}}<$ 4200 GeV. These intervals are fully compatible with the lower mass bounds obtained by the ATLAS and CMS collaborations.

It is possible in a supersymmetric theory to have a stop quark or a stau lepton as the next to lightest supersymmetric particle (NLSP). For such a scenario, we can obtain the desired LSP relic abundance in the universe through co-annihilation.  However, we see from the $m_{\tilde{t}_{1}}-m_{\tilde{\chi}^{0}_1}$, $m_{\tilde{\tau}_{1}}-m_{\tilde{\chi}^{0}_1}$ panels that the maroon points are far from the unit line. This  indicates that in the 4-2-2 model  we cannot have neutralino-stop (or stau) co-annihilation scenario. Therefore, the dark matter relic  abundance  required by  WMAP is not compatible in the 4-2-2 model with the resolution of the little hierarchy problem.

In the  $m_{\tilde{\chi}^{+}_1}-m_{\tilde{\chi}^{0}_1}$ plane we observe that the maroon points  lie near the unit line, which indicates that the
lightest neutralino  is mostly higgsino-like. This yields relatively low values for  relic abundance unless the LSP neutralino mass is around 1 TeV \cite{Baer:2006te}.
If we relax the condition for natural SUSY and look for solutions with $\Delta_{EW}<150$ and $\Delta_{HS}<150$, the higgsino-like LSP neutralino can be the desired dark matter candidate.

Motivated by the recent results from Higgs searches at the LHC, we also calculate the Higgs production and decay in the diphoton channel for the 4-2-2 model.
 The particle spectrum was calculated using the ISAJET~7.84 package~\cite{ISAJET} and this was interfaced with FeynHiggs 2.9.4~\cite{feynhiggs} to estimate the production cross section and decay width.
 We introduce a parameter $R_{\gamma \gamma}$ to quantify possible excess in Higgs production and decay in the diphoton channel  over the Standard Model expectation,
 \begin{eqnarray}
R_{\gamma \gamma} \equiv \frac{\sigma(h) \times Br(h\rightarrow \gamma \gamma)}{(\sigma(h) \times Br(h\rightarrow \gamma \gamma))_{SM}}.
\label{eq:ratio}
\end{eqnarray}
In Figure \ref{fig7} we show the results
in $R_{\gamma \gamma} - \Delta_{EW}$  and $R_{\gamma \gamma} - \Delta_{HS}$ planes for the 4-2-2 model. Color coding
is the same as described in Figure \ref{fig1}.
 We did not find any solution in the natural SUSY limit corresponding to
 $R_{\gamma \gamma}>1.2$. In the   $R_{\gamma \gamma} - \Delta_{EW}$ plane there are some orange points with $R_{\gamma \gamma}>1.2$ with $\Delta_{EW}$ and $\Delta_{HS}$ both larger than 400, but these solution do not satisfy the Higgs mass bound (green points).
 As shown, for instance, in ref. \cite{Ajaib:2012eb},  in order to have $R_{\gamma \gamma}>1.2$ in the MSSM it is necessary to have  either a stop  or stau  around 100 GeV. This condition is not sufficient but, for this discussion, we will restrict ourselves to the necessary condition.  In  Figure \ref{fig6}  from  the  $m_{\tilde{t}_{1}}-m_{\tilde{\chi}^{0}_1}$ and  $m_{\tilde{\tau}_{1}}-m_{\tilde{\chi}^{0}_1}$ planes we observe that the Higgs mass bound (green points) requires minimal values for the stop and stau to be larger than 500 GeV. This explains why we do not find solutions with   $R_{\gamma \gamma}>1.2$  in this model.


In Table 1, we show three benchmark points satisfying the various constraints
mentioned in Section \ref{constraintsSection}. These display the minimal values 
 of $\Delta_{EW}$
and $\Delta_{HS}$  
 that are compatible with  124 GeV CP-even Higgs
boson. Point 1 displays solution with  minimal value of $\Delta_{EW}$
and $\Delta_{HS}$  in the framework of CMSSM. Point 2 represents minimal value of $\Delta_{EW}$
and $\Delta_{HS}$ in the 4-2-2 model. Point 3 depict solutions corresponding minimal $\Delta_{EW}$ and
$\Delta_{HS}$ and best $\Omega_{CDM}h^{2}$ values.


\section{Conclusion \label{conclusions}}

By imposing conditions for natural SUSY  ($\Delta_{EW}<100$ and $\Delta_{HS}<100$)  and   requiring 123 GeV $<m_h<$ 127 GeV,  we obtain a distinctive particle spectra characterized by relatively light third generation sfermions. The light stop quark mass lies in the range 700 GeV $<m_{\tilde{t}_{1}}<$ 1500 GeV, and the range for the light stau lepton mass is  900 GeV $<m_{\tilde{\tau}_{1}}<$ 1300 GeV. The first two family squarks lie in the mass range 3000 GeV $<m_{\tilde{t}_{1}}<$ 4500 GeV, and for the gluino we find 2500 GeV $<m_{\tilde{g}_{1}}<$ 3500 GeV. It is interesting that the bound for $\tan\beta$ is 10 $<\tan\beta<$ 20. Although the higgsino-like chargino and neutralino can be $O(100)$ GeV,  was shown in ref. \cite{Baer:2011ec} observing it at the LHC may still be difficult.
If  the excess in the Higgs production and decay in the diphoton channel is confirmed in future experiments it will rule out this class of Supersymmetric $SU(4)_c \times SU(2)_L \times SU(2)_R$ model with natural Supersymmetry.

\section*{Acknowledgments}
We thank   M. Adeel Ajaib  for valuable discussions.
This work is supported in part by the DOE Grant No. DE-FG02-12ER41808. This work used the Extreme Science
and Engineering Discovery Environment (XSEDE), which is supported by the National Science
Foundation grant number OCI-1053575.

\thispagestyle{empty}


\end{document}